# Engineering topological exciton structures in two-dimensional semiconductors by a periodic electrostatic potential

Na Zhang[1], Wang Yao[2,3], Hongyi Yu[1,4*]

[1] Guangdong Provincial Key Laboratory of Quantum Metrology and Sensing & School of Physics and Astronomy, Sun Yat-Sen University (Zhuhai Campus), Zhuhai 519082, China

[2] New Cornerstone Science Laboratory, Department of Physics, University of Hong Kong, Hong Kong, China

[3] HK Institute of Quantum Science & Technology, University of Hong Kong, Hong Kong, China

[4] State Key Laboratory of Optoelectronic Materials and Technologies, Sun Yat-Sen University (Guangzhou Campus), Guangzhou 510275, China

* E-mail: yuhy33@mail.sysu.edu.cn

**Abstract:** We show the ability of hybridizing different Rydberg states by a periodic electrostatic potential provides a conceptually new scheme for engineering topological exciton structures in layered transition metal dichalcogenides. Such a potential can be remotely imprinted from charge distributions in substrate layers, whose large tunability gives rise to rich topological phase diagrams for the exciton. We find the topological lowest band of the dipolar interlayer exciton can exhibit a small bandwidth, as well as remarkable quantum geometries well suited for realizing the long-sought bosonic fractional Chern insulator. For monolayer excitons, topological bands and in-gap helical edge states can emerge near the energy of 2p states.

## I. Introduction

In the emerging field of two-dimensional (2D) semiconducting transition metal dichalcogenides (TMDs), spin and other quantum degrees-of-freedom (DoF) can greatly affect their electronic, optical and topological properties [1]. The band edges of monolayer and certain bilayer structures of TMDs are located at $\mathbf{K}$ and $\bar{\mathbf{K}}$ corners of the hexagonal Brillouin zone, introducing a valley DoF (or pseudospin) to low-energy carriers. These two valleys are related by a time reversal and locked to the spin at a given energy due to the strong spin-orbit coupling of the transition metal atom [2]. Carriers in $\mathbf{K}$ and $\bar{\mathbf{K}}$ valleys can be distinguished by their opposite Berry curvatures and valley magnetic moments originating from the intrinsic Bloch band geometry [1]. The resultant valley Hall effect under an in-plane electric field [3] and valley splitting under an out-of-plane magnetic field [4-7] allow the tunability of valley DoF through external fields. In a van der Waals bilayer structure formed by two vertically stacked monolayers, a large-scale moiré pattern with spatially varying atomic registries can emerge. The tunable moiré potential provides an excellent platform for investigating various correlated and topological quantum phenomena [8,9]. Carriers in bilayer moiré systems also exhibit a layer DoF, with layer-pseudospin orientations determined by the spatially modulating layer-hybridizations. The resultant precession of the layer-pseudospin with position can give rise to nontrivial topological structures for carriers in $\mathbf{K}$ (or $\bar{\mathbf{K}}$) valley [10,11], which is at the heart of the recently discovered integer/fractional Chern insulators in bilayer TMDs moiré patterns [12-18].

The exciton plays a crucial role in optoelectronic properties of layered TMDs [19-21]. Its motion consists of a center-of-mass (CoM) part, and an electron-hole (e-h) relative part described by the discrete Rydberg series 1s, 2s, $2p_{\pm}$, … [22-25]. A large portion of previous works focus on the lowest-energy 1s state, and study how those DoFs inherited from the electron/hole constituent affect the exciton CoM motion [26-33]. For the intravalley spin-singlet bright exciton in monolayer TMDs, its two valleys are efficiently coupled by the e-h exchange interaction, resulting in a momentum-dependent valley-orbit coupling [19,34-36]. Meanwhile, layer-hybridized excitons with tunable layer DoF have been detected in bilayer TMDs [37-40]. The variation of the excitonic valley/layer DoF with momentum/position in a moiré pattern can bring nontrivial topological exciton structures. Combined with the long lifetime and strong dipolar interactions, the topological interlayer exciton ($X_{inter}$) can serve as an excellent candidate for realizing the bosonic version of the recently observed fermionic fractional Chern insulators (FCI) [12-18], with the out-of-plane electric dipole facilitating its detection through transport measurements. Several schemes have been proposed to realize topological exciton states in TMDs moiré patterns, making use either the e-h exchange induced valley-orbit coupling [41] or the spatially modulating layer-hybridization of the hole constituent [42], or a combination of the two [43].

The Rydberg series can serve as a quantum DoF unique to the exciton with large tunability. Experiments have demonstrated that couplings between 1s and $2p_{\pm}$ (2s and $2p_{\pm}$) states can be introduced by an externally applied infrared optical field [44-46] (in-plane static electric field [47]). Meanwhile, the intrinsic Bloch band Berry curvature of TMDs can lift the degeneracy of $2p_{\pm}$ excitons [48-50], introducing an energy splitting as large as ~ 15 meV [46]. For excitons in TMDs moiré patterns, earlier works mostly focus on how the moiré potential affects the CoM motion, whereas the influence on the Rydberg states has just been noticed. Recent experiments have shown that the optical spectrum near 2s state in monolayer TMDs can be remotely changed by doping moiré-patterned layers in the substrate [51,52], which is attributed to a periodic electrostatic potential remotely imprinted by the charge distribution in the substrate. Such an electrostatic potential enables momentum-dependent hybridizations between different Rydberg states [53,54], thus can serve as a realistic and tunable way to manipulate the Rydberg DoF. However, whether a nontrivial topology can be introduced to the exciton by manipulating its Rydberg DoF is still largely unexplored.

In this work, we propose a conceptually new scheme for engineering topological exciton structures in layered TMDs, making use of the hybridization between Rydberg states induced by a periodic electrostatic potential. We show that, tuning the potential strength and wavelength gives rise to rich topological phase diagrams for the exciton. Remarkably, the topologically nontrivial lowest band of $X_{inter}$ can feature a small bandwidth, as well as quantum geometries well suited for realizing the long-sought bosonic FCI. For monolayer excitons ($X_{mono}$), topological bands and in-gap helical edge states can emerge near the energy of $2p_{\pm}$ states. Our scheme thus offers an excellent platform for exploring novel exciton-related topological phenomena.

## II. Model for excitons in a periodic electrostatic potential

Electrons and holes in layered TMDs are susceptible to external perturbations due to the atomically thin structure. The spatially-modulating charge density in moiré-patterned twisted bilayer graphene (TBG), hexagonal boron nitride (hBN) or TMDs can generate periodic electrostatic potentials to carriers in adjacent TMDs layers [51,52,55-57], see Fig. 1(a). It should be emphasized that although Fig. 1(a) uses an $X_{inter}$ as an illustration, our model can apply to both $X_{inter}$ in bilayer TMDs and $X_{mono}$ in monolayer/bilayer TMDs. We write the potential experienced by the electron (hole) as $U(\mathbf{r_e})$ ($U'(\mathbf{r_h})$), with $\mathbf{r_{e/h}}$ the spatial coordinate of the electron/hole. Fig. 1(b) shows the landscape of $U(\mathbf{r_e})$ induced by a triangular-type spatial charge distribution, forming a honeycomb or triangular lattice with a wavelength $\lambda$. The exciton Hamiltonian can be written as $\hat{H} = \hat{H}_{\mathrm{X}} + \delta\hat{H}_{\mathrm{X}} + \hat{U}_{\mathrm{X}}$. Here $\hat{H}_{\mathrm{X}} = -\frac{\hbar^2}{2m_{\mathrm{e}}}\frac{\partial^2}{\partial \mathbf{r}_{\mathrm{e}}^2} - \frac{\hbar^2}{2m_{\mathrm{h}}}\frac{\partial^2}{\partial \mathbf{r}_{\mathrm{h}}^2} + V(\mathbf{r}_{\mathbf{e}} - \mathbf{r}_{\mathbf{h}})$ is the 2D hydrogen-like free exciton Hamiltonian with $V(\mathbf{r}_{\mathbf{e}} - \mathbf{r}_{\mathbf{h}})$ the Coulomb potential and $m_{\mathbf{e/h}} \approx \mathbf{0.5}m_0$ the electron/hole effective mass ($m_0$ is the free electron mass). Introducing the exciton mass $M = m_{\mathbf{e}} + m_{\mathbf{h}}$, the CoM coordinate $\mathbf{R} \equiv \frac{m_{\mathbf{e}}}{M}\mathbf{r}_{\mathbf{e}} + \frac{m_{\mathbf{h}}}{M}\mathbf{r}_{\mathbf{h}}$ and the relative coordinate $\mathbf{r} \equiv \mathbf{r}_{\mathbf{e}} - \mathbf{r}_{\mathbf{h}}$, the eigenstate $|\mathbf{k}, n\rangle$ of $\hat{H}_{\mathrm{X}}$ can be decomposed into a CoM part $|\mathbf{k}\rangle$ in the plane-wave form and a relative part $|n\rangle$ in the Rydberg series. Here $\mathbf{k}$ is the CoM momentum and $n =$ 1s, 2s, 2p$_\pm$, 3d$_\pm$, …. $\delta\hat{H}_X = -\Omega_{\mathbf{K}/\overline{\mathbf{K}}}\left(\frac{\partial V(\mathbf{r})}{\partial \mathbf{r}} \times i\frac{\partial}{\partial \mathbf{r}}\right)_z$ accounts for the effect of the intrinsic Bloch band Berry curvature $\Omega_{\mathbf{K}} = -\Omega_{\overline{\mathbf{K}}}$ in $\mathbf{K}$ and $\overline{\mathbf{K}}$ valleys, whose major effect is an energy correction to the Rydberg state with a sign determined by its angular momentum [48-50]. Most importantly, it introduces a finite splitting between 2p$_+$ and 2p$_-$ (as well as between 3d$_+$ and 3d$_-$) with opposite signs in $\mathbf{K}$ and $\overline{\mathbf{K}}$ valleys [46,48-50], which is essential as it breaks the inversion symmetry and allows nontrivial topological structures to emerge. The dispersion curves of $|\mathbf{k}, n\rangle$, given by $\frac{\hbar^2 k^2}{2M} + E_n$, are schematically shown in Fig. 1(c). $\hat{U}_{\mathrm{X}} = U(\mathbf{r}_{\mathbf{e}}) + U'(\mathbf{r}_{\mathbf{h}})$ is the total electrostatic potential experienced by the exciton. Note that $U(\mathbf{r}_{\mathbf{e}})$ and $U'(\mathbf{r}_{\mathbf{h}})$ vary smoothly in a moiré wavelength $\lambda \sim 10$ nm, much larger than the length scale of $\frac{m_{\mathrm{e/h}}}{M}\mathbf{r}$ ($\sim \frac{a_{\mathrm{B}}}{2}$ with the Bohr radius $a_{\mathrm{B}} \approx$ 1.7 nm for 1s and 6.6 nm for 2s [58]). We thus adopt a linear expansion on $\frac{m_{\mathrm{e/h}}}{M}\mathbf{r}$ to write $\hat{U}_X \approx \delta U(\mathbf{R}) + \hat{U}_{\mathrm{r}}$, with

$$\begin{aligned} \delta U(\mathbf{R}) &\equiv U(\mathbf{R}) + U'(\mathbf{R}), \\ \hat{U}_{\mathrm{r}} &\equiv \mathbf{r} \cdot \left[\frac{m_{\mathrm{h}}}{M}\nabla U(\mathbf{R}) - \frac{m_{\mathrm{e}}}{M}\nabla U'(\mathbf{R})\right]. \end{aligned} \tag{1}$$

The periodic potential $\delta U(\mathbf{R})$ only affects $|\mathbf{k}\rangle$, and the resultant formation of moiré bands for the exciton CoM motion has been widely studied in previous works [26-33]. Meanwhile, $\hat{U}_{\mathrm{r}}$ affects both the CoM and relative motions, which couples $|\mathbf{k}, n\rangle$ and $|\mathbf{k}', n'\rangle$ with $\mathbf{k} \neq \mathbf{k}'$ and $n \neq n'$. As a result, the exciton eigenstates under $\hat{U}_{\mathrm{r}}$ show momentum-dependent hybridizations between different Rydberg states [53], which can be used to implement nontrivial topologies to excitons. We do not consider the e-h exchange interaction until in the last section.

Before presenting quantitative results, we shall first use a simple toy model to give some qualitative insights for the emergence of nontrivial topologies. Without considering $\hat{U}_{\mathrm{r}}$, $\delta U(\mathbf{R})$ gives rise to decoupled moiré bands for different Rydberg states which only differ by constant energy shifts. We focus on bands of the two lowest-energy Rydberg states 1s and 2p$_-$ of the $\mathbf{K}$ valley $\mathrm{X}_{\mathrm{inter}}$, see Fig. 1(d). For suitable strengths of $\delta U(\mathbf{R})$, the second band of 1s and the lowest one of 2p$_-$ can cross in the vicinity of $\mathbf{\Gamma}$, whose corresponding states at $\mathbf{\Gamma}$ are denoted as $\mathbf{\Gamma}_{\mathrm{s2}}$, and $\mathbf{\Gamma}_{\mathrm{p1}}$, respectively (Fig. 1(e)). Given the $2\pi/3$-rotational ($\hat{C}_3$) symmetry of the potentials, excitons at $\mathbf{\Gamma}$ are characterized by a quantum number $C_{3\mathrm{X}} = 0, \pm 1$ which is the total $\hat{C}_3$ quantum number of CoM and relative motions. Note that $C_{3\mathrm{X}}(\mathbf{\Gamma}_{\mathrm{s2}})$ is determined by $\delta U(\mathbf{R})$ and can be different from $C_{3\mathrm{X}}(\mathbf{\Gamma}_{\mathrm{p1}}) = -1$. The effect of $\hat{U}_{\mathrm{r}}$ is to introduce a momentum-dependent coupling between the two bands, with a form $\propto k e^{\pm i\theta_{\mathbf{k}}}$ when $C_{3\mathrm{X}}(\mathbf{\Gamma}_{\mathrm{p1}}) - C_{3\mathrm{X}}(\mathbf{\Gamma}_{\mathrm{s2}}) = \pm 1$ for $\mathbf{k}$ near $\mathbf{\Gamma}$. Here $\theta_{\mathbf{k}}$ is the direction angle of $\mathbf{k}$. This then opens a gap and introduces topological band inversions with Chern numbers $\mathcal{C} = \pm 1$, see Fig. 1(e). Such an analysis only serves as an intuitive picture for certain parameter regimes. For general cases, realistic models are needed for quantitative results.

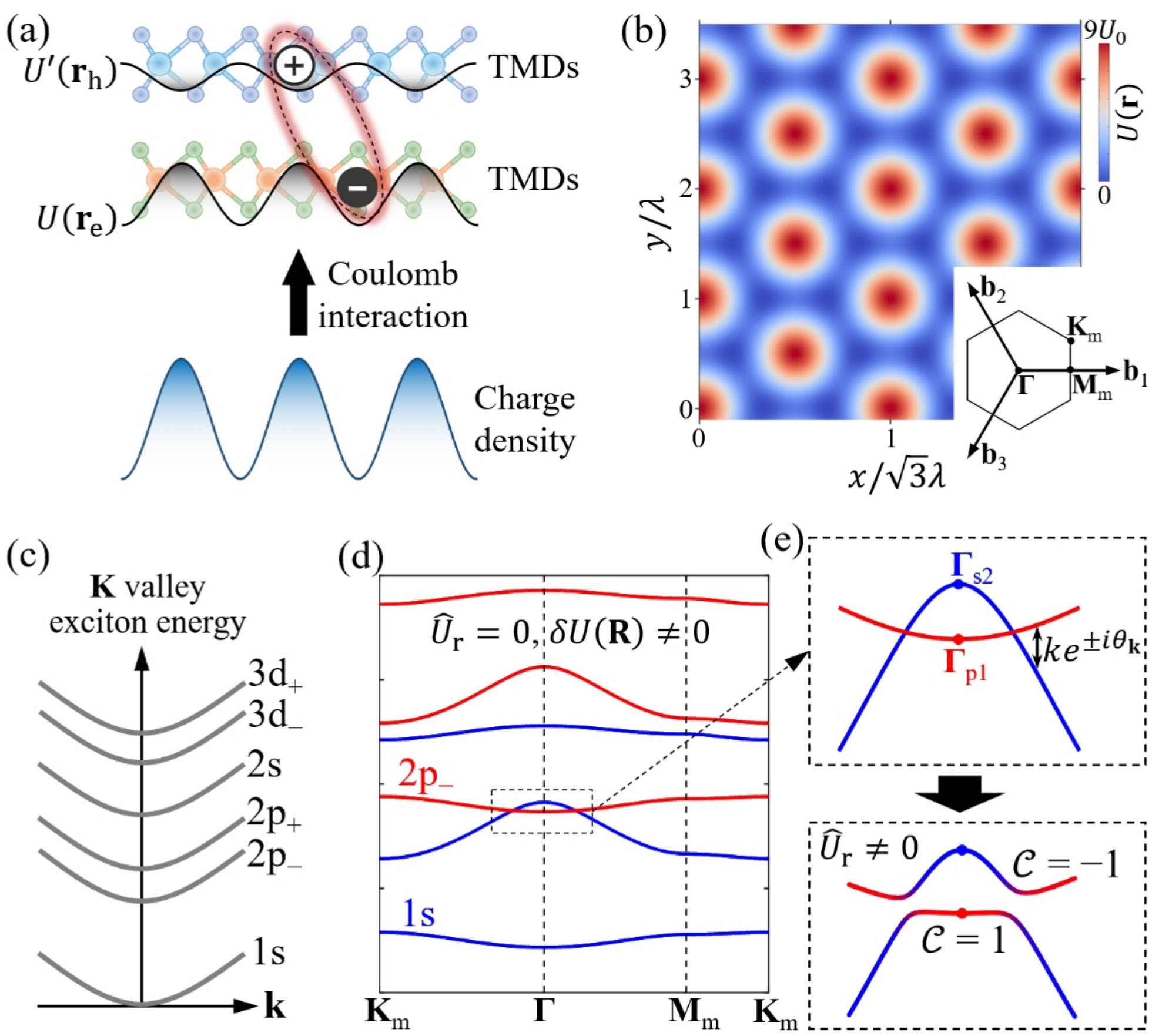

Figure 1. (a) A schematic illustration of the model, where an exciton in layered TMDs is subjected to a periodic electrostatic potential remotely generated by the proximity charge distribution. The potentials experienced by the electron and hole are denoted as $U(\mathbf{r}_{\mathrm{e}})$ and $U'(\mathbf{r}_{\mathrm{h}})$, respectively. (b) The landscape of $U(\mathbf{r})$ induced by a triangular-type charge distribution, forming a honeycomb (triangular) lattice when $U_0 > 0$ ($U_0 < 0$). The inset shows the moiré Brillouin zone, with $\pm\mathbf{b}_{1,2,3}$ the smallest nonzero moiré reciprocal lattice vectors. (c) Dispersion curves of Rydberg states for **K** valley free excitons without considering the e-h exchange interaction. (d) A schematic picture for moiré bands of 1s and 2p− states under the effect of $\delta U(\mathbf{R})$. The lowest band of 2p− and second-lowest one of 1s can cross near $\mathbf{\Gamma}$ for suitable $\delta U(\mathbf{R})$ strengths. (e) A toy model for the $\hat{U}_{\mathrm{r}}$-induced topological exciton structure. Near $\mathbf{\Gamma}$, $\hat{U}_{\mathrm{r}}$ introduces a coupling $\propto ke^{\pm i\theta_{\mathbf{k}}}$ to the two bands when excitons at $\mathbf{\Gamma}_{\mathrm{p1}}$ and $\mathbf{\Gamma}_{\mathrm{s2}}$ have different $\hat{C}_3$ quantum numbers. This then introduces topological band inversions and nonzero Chern numbers.

The periodic potential can be expanded into the Fourier series $U(\mathbf{R}) = \sum_{\mathbf{G}} e^{i\mathbf{G}\cdot\mathbf{R}} U(\mathbf{G})$ with $\mathbf{G}$ the moiré reciprocal lattice vector. Given that $U(\mathbf{R})$ and $U'(\mathbf{R})$ are smooth functions of $\mathbf{R}$, their Fourier coefficients $U(\mathbf{G})$ and $U'(\mathbf{G})$ should decrease rapidly with $|\mathbf{G}|$. Generally it is sufficient to keep only $\mathbf{G} = \pm\mathbf{b}_{1,2,3}$ which correspond to the nonzero moiré reciprocal lattice vectors with the smallest magnitudes (see Fig. 1(b) inset). The matrix element of $\hat{U}_{\mathrm{r}}$ is $\langle \mathbf{k}, n | \hat{U}_{\mathrm{r}} | \mathbf{k}', n' \rangle = \sum_{\mathbf{G}\neq 0} \delta_{\mathbf{k}-\mathbf{k}',\mathbf{G}} t_{nn'}(\mathbf{G})$ with $t_{nn'}(\mathbf{G}) \equiv i\left[\frac{m_{\mathrm{h}}}{M} U(\mathbf{G}) + \frac{m_{\mathrm{e}}}{M} U'(\mathbf{G})\right] \mathbf{G} \cdot \langle n | \hat{\mathbf{r}} | n' \rangle$. Note that $\langle n | \hat{\mathbf{r}} | n' \rangle \neq 0$ only if $|n\rangle$ and $|n'\rangle$ have an angular momentum difference of ±1. For the low-energy Rydberg states 1s, $2\mathrm{p}_{\pm}$, 2s and $3\mathrm{d}_{\pm}$, only $\langle 2\mathrm{p}_{\pm} | \hat{\mathbf{r}} | 1\mathrm{s} \rangle$, $\langle 2\mathrm{p}_{\pm} | \hat{\mathbf{r}} | 2\mathrm{s} \rangle$ and $\langle 2\mathrm{p}_{\pm} | \hat{\mathbf{r}} | 3\mathrm{d}_{\pm} \rangle$ are finite. For a momentum **k** restricted to the moiré Brillouin zone (mBZ), the **K** valley exciton Hamiltonian becomes $\hat{H} = \sum_{\mathbf{k}\in\mathrm{mBZ}} \hat{H}_{\mathbf{k}}$ with

$$
\begin{aligned}
\hat{H}_{\mathbf{k}} = & \sum_{n}\sum_{\mathbf{G}}\left(\frac{\hbar^2(\mathbf{k}+\mathbf{G})^2}{2M}+E_n\right)|\mathbf{k}+\mathbf{G},n\rangle\langle\mathbf{k}+\mathbf{G},n| \\
& +\sum_{j=1}^{3}\sum_{n}\left(\delta U(\mathbf{b}_j)\sum_{\mathbf{G}}|\mathbf{k}+\mathbf{G}+\mathbf{b}_j,n\rangle\langle\mathbf{k}+\mathbf{G},n|+\mathrm{h.c.}\right) \\
& +\sum_{j=1}^{3}\sum_{nn'}\left(t_{nn'}(\mathbf{b}_j)\sum_{\mathbf{G}}|\mathbf{k}+\mathbf{G}+\mathbf{b}_j,n\rangle\langle\mathbf{k}+\mathbf{G},n'|+\mathrm{h.c.}\right).
\end{aligned} \tag{2}
$$

Here $\delta U(\mathbf{b}_j) \equiv U(\mathbf{b}_j) + U'(\mathbf{b}_j)$. The Hamiltonian of Eq. (2) can be diagonalized to get the eigenstate $|\psi_{n_X,\mathbf{k}}\rangle = \sum_{\mathbf{G},n}\langle\mathbf{k}+\mathbf{G},n|\psi_{n_X,\mathbf{k}}\rangle|\mathbf{k}+\mathbf{G},n\rangle$ of $n_X$-th exciton band, which now becomes the hybridization of different Rydberg states. Such a hybrid wave function is used to calculate the Berry curvature $\Omega_{n_X,\mathbf{k}} \equiv i\langle\nabla_{\mathbf{k}}u_{n_X,\mathbf{k}}| \times |\nabla_{\mathbf{k}}u_{n_X,\mathbf{k}}\rangle\cdot\mathbf{z}$ and the Chern number $C_{n_X} = \frac{1}{2\pi}\int_{\mathrm{mBZ}} d\mathbf{k}\Omega_{n_X,\mathbf{k}}$, where $u_{n_X,\mathbf{k}}$ is the periodic part of $\psi_{n_X,\mathbf{k}}$.

Theoretical and experimental works have indicated that, the alternating ferroelectric domains in a twisted hBN substrate can impose an electrostatic potential $U(\mathbf{R})$ with an $O(0.1)$ eV modulation range to the adjacent TMDs monolayer [56,57], resulting in $|U(\mathbf{b}_j)|$ in the order of several tens meV. Meanwhile, the strength of $t_{nn'}(\mathbf{b}_j)$ is proportional to $\langle n|\hat{\mathbf{r}}|n'\rangle$. Experiments have obtained $|\langle 2\mathrm{p}_\pm|\hat{\mathbf{r}}|1\mathrm{s}\rangle| \approx$ 1 nm in monolayer $MoSe_2$ [46], whereas numerical calculations give $|\langle 2\mathrm{p}_\pm|\hat{\mathbf{r}}|2\mathrm{s}\rangle| \approx$ 4 nm which is three to four times larger than $|\langle 2\mathrm{p}_\pm|\hat{\mathbf{r}}|1\mathrm{s}\rangle|$ but slightly smaller than $|\langle 2\mathrm{p}_\pm|\hat{\mathbf{r}}|3\mathrm{d}_\pm\rangle|$ [59]. The resultant inter-Rydberg coupling strengths $t \equiv |t_{2\mathrm{p}_\pm,2\mathrm{s}}(\mathbf{b}_j)| \sim 3|t_{2\mathrm{p}_\pm,1\mathrm{s}}(\mathbf{b}_j)| \sim |t_{2\mathrm{p}_\pm,3\mathrm{d}_\pm}(\mathbf{b}_j)|$ are estimated to be several tens meV. Furthermore, $t$ can be continuously tuned from 0 to several tens meV when the electrostatic potential comes from the doped charge distributions in adjacent moiré-patterned TBG or TMDs layers, see Appendix A for details. Note that $U'(\mathbf{R}) = -U(\mathbf{R})$ for $X_{\mathrm{mono}}$ with the electron and hole in the same layer, but $U'(\mathbf{R}) \neq -U(\mathbf{R})$ otherwise. Below we treat $\delta U_0 \equiv \delta U(\mathbf{b}_j)$ and $t$ of $X_{\mathrm{inter}}$ as independent parameters, considering that $U(\mathbf{r}_e)$ ($U'(\mathbf{r}_h)$) in the lower (upper) TMDs layer can be tuned sensitively by charge distributions in the bottom substrate (top capping layer).

## III. Topological flat bands and quantum geometries of interlayer excitons

For $X_{\mathrm{inter}}$ whose Rydberg states are separated by $O(10)$ meV, we set $E_{1\mathrm{s}} = 0$, $E_{2\mathrm{p}_-} = 25$ meV, $E_{2\mathrm{p}_+} = 40$ meV, $E_{2\mathrm{s}} = 60$ meV, $E_{3\mathrm{d}_-} = 75$ meV, $E_{3\mathrm{d}_+} = 85$ meV, and fix the ratios between inter-Rydberg coupling strengths at $t \equiv |t_{2\mathrm{p}_\pm,2\mathrm{s}}(\mathbf{b}_j)| = 3|t_{2\mathrm{p}_\pm,1\mathrm{s}}(\mathbf{b}_j)| = \frac{3}{4}|t_{2\mathrm{p}_\pm,3\mathrm{d}_\pm}(\mathbf{b}_j)|$. Fig. 2(a) shows the calculated **K** valley $X_{\mathrm{inter}}$ bands under $\lambda = 10$ nm, $\delta U_0 = 30$ meV and $t = 25$ meV. The obtained three lowest bands are isolated from each other with global gaps in the order of 1 meV, whose Chern numbers are $(C_1, C_2, C_3) = (1,-1,0)$. The $\bar{\mathbf{K}}$ valley $X_{\mathrm{inter}}$ exhibits opposite Chern numbers due to the time-reversal symmetry. The small bandwidth of the lowest band (smaller than the global gap $\Delta_{12}$ between the two lowest bands) implies that it can serve as a candidate system for studying the bosonic FCI. We note that besides the nearly flat condition, the stability of the FCI has additional constraints on quantum geometries of the topologically nontrivial band [60,61]. To be specific, the Berry curvature $\Omega_{1,\mathbf{k}}$ should be largely homogeneous in the entire mBZ, and the non-negative quantity $T_{1,\mathbf{k}} \equiv \mathrm{Tr}(g_{1,\mathbf{k}}) - |\Omega_{1,\mathbf{k}}|$ (the trace condition) should be close to 0. Here the 2x2 matrix $g_{1,\mathbf{k}}$

corresponds to the quantum metric of the lowest band, with matrix elements given by $g_{1,\mathbf{k}}^{\mu\nu} \equiv \mathbf{Re}\left\langle \frac{\partial u_{1,\mathbf{k}}}{\partial k_\mu} \right| \left(1-|u_{1,\mathbf{k}}\rangle\langle u_{1,\mathbf{k}}|\right) \left| \frac{\partial u_{1,\mathbf{k}}}{\partial k_\nu}\right\rangle$ for $\mu,\nu = x,y$. We show the distributions of $\Omega_{1,\mathbf{k}}/\bar{\Omega}_1$ and $T_{1,\mathbf{k}}/\bar{\Omega}_1$ in Fig. 2(b), where $\bar{\Omega}_1 = 2\pi\mathcal{C}_1/A_{\mathrm{mBZ}}$ is the average Berry curvature with $A_{\mathrm{mBZ}}$ the mBZ area. The inhomogeneity of $\Omega_{1,\mathbf{k}}$ and the deviation of $T_{1,\mathbf{k}}$ from 0 over the entire mBZ can be quantified by the dimensionless Berry curvature fluctuation $\sigma_\Omega \equiv \frac{1}{2\pi}\sqrt{A_{\mathrm{mBZ}}\int_{\mathrm{mBZ}} d\mathbf{k}\left(\Omega_{1,\mathbf{k}}^2-\bar{\Omega}_1^2\right)}$ and trace condition violation $\sigma_{\mathrm{T}} \equiv \frac{1}{2\pi}\int_{\mathrm{mBZ}} d\mathbf{k}T_{1,\mathbf{k}}$, respectively. In Fig. 2(b), $\sigma_\Omega = 0.136$ and $\sigma_{\mathrm{T}} = 0.126$ are indeed close to 0.

The values of $\Delta_{12}$, $\mathcal{C}_1$, $\sigma_\Omega$ and $\sigma_{\mathrm{T}}$ can all be tuned by the periodic potentials. Fig. 2(c) is a 2D map of $\Delta_{12}$ in $(\delta U_0, t)$ parameter space under $\lambda = 10$ nm. Here we only show the $\delta U_0 \geq 0$ case which corresponds to a honeycomb potential (see Fig. 1(b)) experienced by the $\mathrm{X_{inter}}$ CoM motion, whereas $\delta U_0 < 0$ is found to result in a trivial topology. The different gapped ($\Delta_{12} > 0$) regimes with various $\mathcal{C}_1$ values are separated by gapless ($\Delta_{12} = 0$) boundaries, showing a rich topological phase diagram for the lowest $\mathrm{X_{inter}}$ band. Fig. 2(d) and 2(e) give the values of $\sigma_\Omega$ and $\sigma_{\mathrm{T}}$, respectively. As can be seen, both $\sigma_\Omega$ and $\sigma_{\mathrm{T}}$ are much smaller than 1 (especially $\sigma_{\mathrm{T}}$) in a large area of the $\mathcal{C}_1 = 1$ topological regime. Combined with the out-of-plane electric dipole which provides strong dipolar interactions [62] as well as electric signals in transport measurements, the above properties suggest that the $\mathrm{X_{inter}}$ modulated by a periodic electrostatic potential can be an excellent platform for realizing the long-sought bosonic FCI.

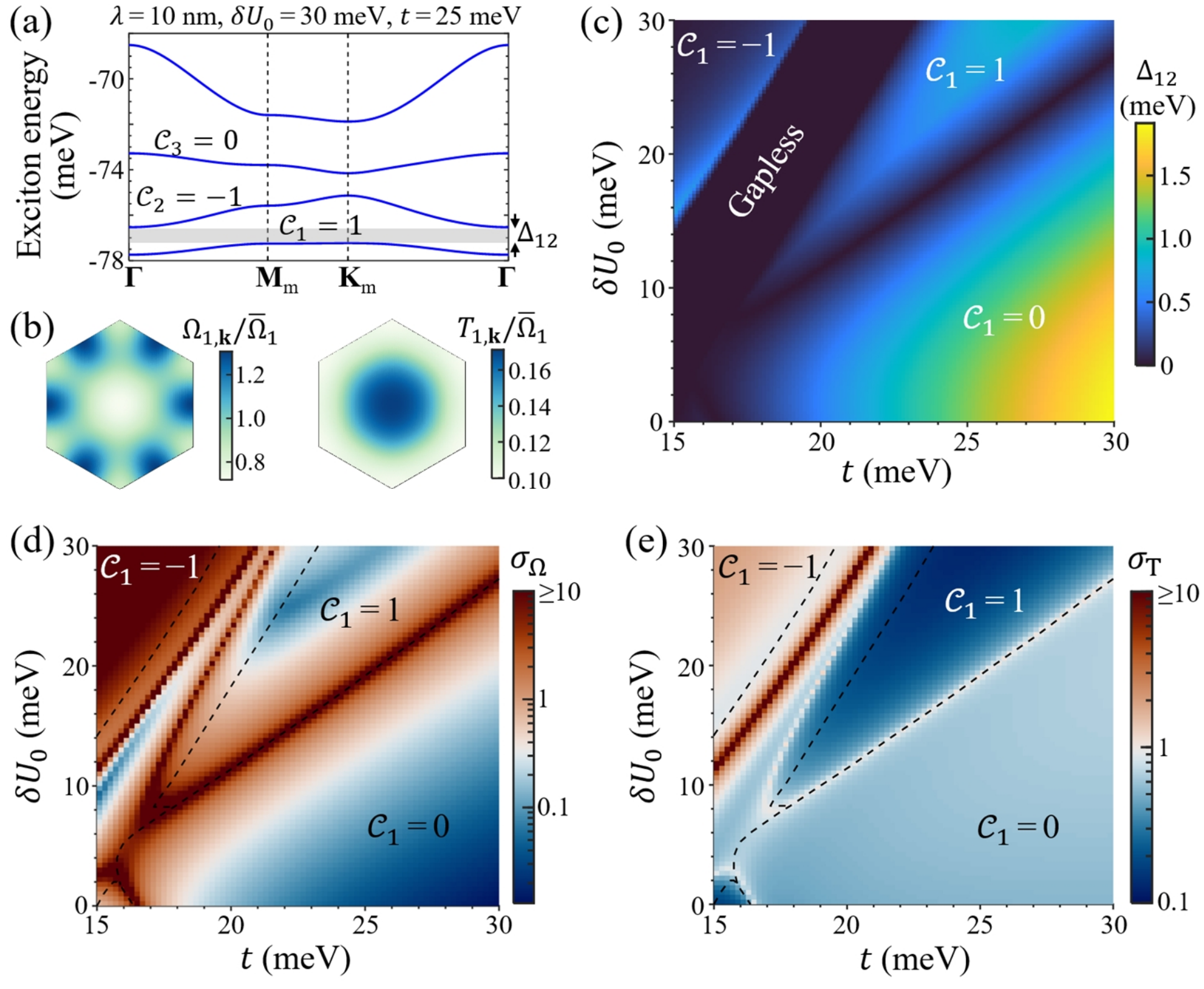

Figure 2. (a) The band structure of the $\mathbf{K}$ valley $\mathrm{X_{inter}}$ under $\lambda = 10$ nm, $\delta U_0 = 30$ meV and $t = 25$ meV. The three lowest bands have Chern numbers $(\mathcal{C}_1, \mathcal{C}_2, \mathcal{C}_3) = (1, -1, 0)$. $\Delta_{12}$ is the global gap between the two lowest bands. (b) The distributions of the Berry curvature $\Omega_{1,\mathbf{k}}$ and trace condition $T_{1,\mathbf{k}}$ in mBZ for the lowest band in (a). (c) $\Delta_{12}$ as a function of $\delta U_0$ and $t$ under $\lambda = 10$ nm. Gapped regimes ($\Delta_{12} > 0$) with different Chern numbers are separated by boundaries with $\Delta_{12} = 0$. (d) The dimensionless Berry curvature fluctuation $\sigma_\Omega$ and (e) trace condition violation $\sigma_\mathrm{T}$ of the lowest band in $(\delta U_0, t)$ parameter space. Dashed lines mark the boundaries of different topological phases.

A small bandwidth can enhance the effect of the exciton-exciton interaction. Therefore, we wish to explore whether an even higher flatness of the lowest topological band can be achieved by tuning the potentials. Fig. 3(a) shows the flatness ratio (the ratio between the gap $\Delta_{12}$ and the bandwidth of the lowest band) as a function of $\delta U_0$ and $t$ under $\lambda$ = 10 nm. Trivial flat bands with flatness ratios ~ 30 appear in a wide parameter regime with small $\delta U_0$ but large $t$ values. Meanwhile, topologically nontrivial flat bands can be found in a narrow parameter regime, with flatness ratios reaching 20. Fig. 3(b) gives the **K** valley $\mathrm{X}_{\mathrm{inter}}$ bands under $\lambda$ = 10 nm and $\delta U_0$ = $t$ = 25 meV, where the lowest band has $\mathcal{C}_1$ = 1 and flatness ratio ≈ 22. We note that the band dispersion of a general model is fundamentally independent on the quantum geometry determined by the wave function. Therefore, topological flat bands do not implicitly have small Berry curvature fluctuations and trace condition violations. For the topological flat bands with large flatness ratios in Fig. 3(a,b), we find they still exhibit quantum geometries good for realizing the FCI, with $\sigma_\Omega \approx$ 0.4 and $\sigma_\mathrm{T} \approx$ 0.2.

Besides $\delta U_0$ and $t$, the moiré wavelength $\lambda$ is another tunable parameter that can largely affect the topological property and band flatness. Fig. 3(c) gives a topological phase diagram of the lowest $\mathrm{X}_{\mathrm{inter}}$ band in $(t, \lambda)$ parameter space under $\delta U_0$ = 25 meV, showing that smaller $\lambda$ values facilitate larger topological gaps at a cost of requiring larger $t$ values. Fig. 3(d) shows the flatness ratio of the lowest $\mathrm{X}_{\mathrm{inter}}$ band under $\delta U_0$ = 25 meV. The narrow regimes with $\mathcal{C}_1$ = 1 and flatness ratios ~ 20 in Fig. 3(a,d) imply there could be some quantum interference effect. Further investigations are needed to understand the underlying mechanism of these topological flat bands.

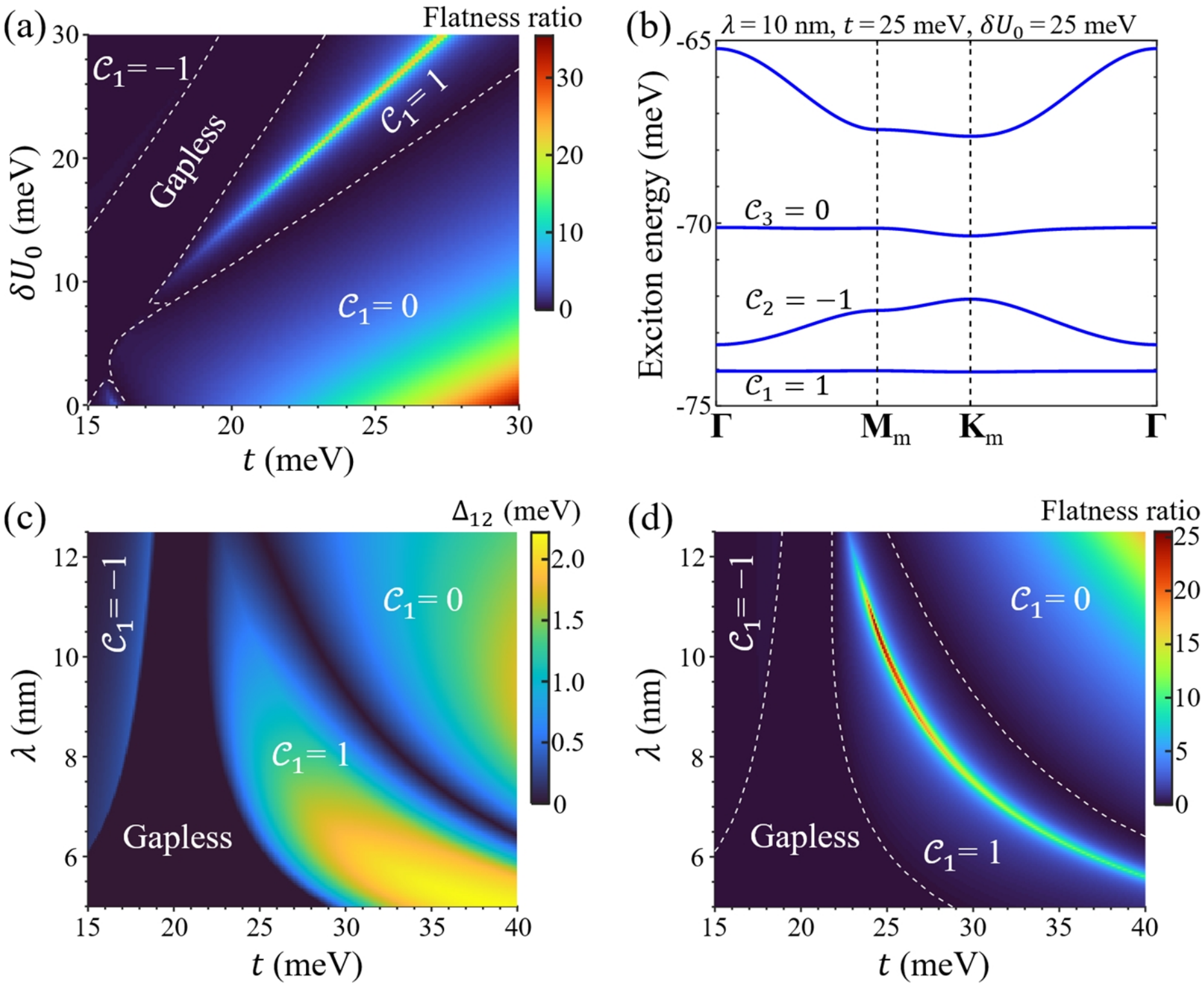

Figure 3. (a) The flatness ratio of the lowest band as a function of $\delta U_0$ and $t$ under $\lambda$ = 10 nm. (b)

The **K** valley $X_{inter}$ bands under $\lambda$ = 10 nm, $\delta U_0$ = 25 meV and $t$ = 25 meV. Chern numbers of the three lowest bands are $(\mathcal{C}_1, \mathcal{C}_2, \mathcal{C}_3) = (1, -1, 0)$. (c) The global gap $\Delta_{12}$ and (d) flatness ratio of the lowest band as functions of $t$ and $\lambda$ under $\delta U_0$ = 25 meV.

To better understand how different wavelengths affect the topological exciton properties, we summarize the values of $\Delta_{12}$, $\sigma_\Omega$, $\sigma_\mathrm{T}$ and flatness ratio of the lowest $X_{inter}$ band as functions of $(\delta U_0, t)$ in Fig. 4, under wavelengths of $\lambda$ = 5, 7, 12 nm and with other parameters the same as in Fig. 2. Similar to the previously considered $\lambda$ = 10 nm case, all wavelengths show the emergence of topologically nontrivial regimes with $\mathcal{C}_1$ = 1, small bandwidths and $\sigma_\Omega, \sigma_\mathrm{T} \ll 1$, which are well suited for realizing the bosonic FCI. Meanwhile, the area of the $\mathcal{C}_1$ = 1 regime in $(\delta U_0, t)$ parameter space becomes larger under smaller $\lambda$ values. Notice that under a sufficiently small $\lambda$ value (e.g., 5 nm in Fig. 4), the topologically nontrivial regime becomes insensitive to $\delta U_0$, but mainly determined by $t$.

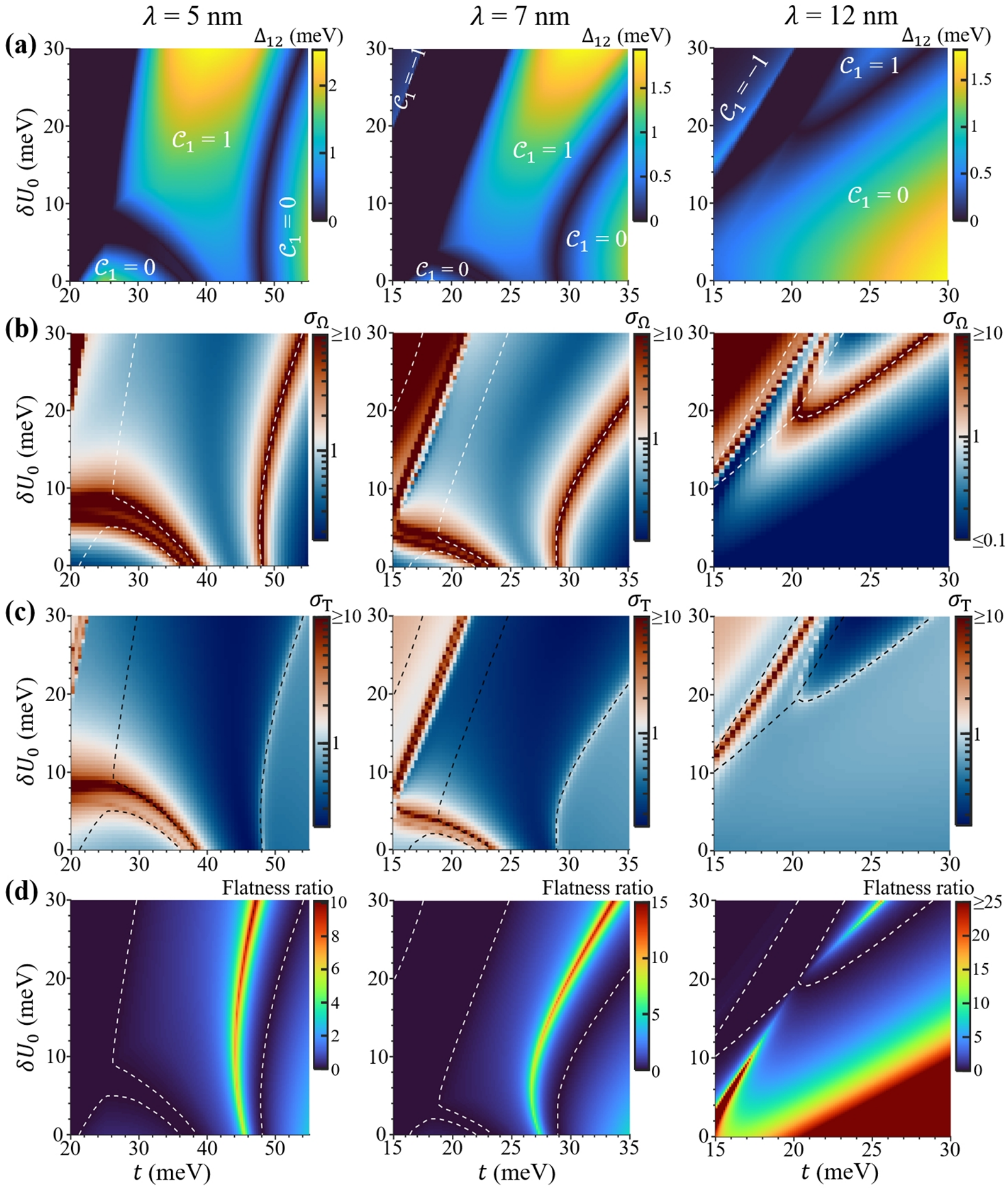

Figure 4. The (a) global gap value $\Delta_{12}$, (b) dimensionless Berry curvature fluctuation $\sigma_\Omega$, (c) trace condition violation $\sigma_\mathrm{T}$, and (d) flatness ratio of the lowest interlayer exciton band in $(\delta U_0, t)$ parameter space under $\lambda$ = 5, 7, and 12 nm. The other parameters are the same as in Fig. 2. Dashed lines mark the boundaries (with $\Delta_{12}$ = 0) of different topological phase regimes.

## IV. Topological monolayer excitons with and without e-h exchange interaction

Below we turn to $X_{mono}$ with $\delta U(\mathbf{R}) = 0$ and $\hat{U}_r \neq 0$. Considering that $E_{2p_\pm} - E_{1s} \sim 130$ meV [44,46] in monolayer TMDs is an order of magnitude larger than $|t_{2p_\pm,1s}(\mathbf{b}_j)| \sim 10$ meV, the hybridization between 1s and $2p_\pm$ is expected to be weak. Meanwhile, the separation between $2p_\pm$ and $3d_\pm$ of $X_{mono}$ is also significantly larger than $X_{inter}$ [59]. To simplify the calculation, below we retain only 2s and $2p_\pm$ states and fix $E_{2p_-} = 0$, $E_{2p_+} = 15$ meV, $E_{2s} = 60$ meV. Fig. 5(a) shows the calculated **K** valley $X_{mono}$ bands under $\lambda = 7$ nm and $t = 20$ meV. The three lowest bands are isolated from each other with global gaps in the order of meV and Chern numbers $(\mathcal{C}_1, \mathcal{C}_2, \mathcal{C}_3) = (1, -1, 0)$. By numerically solving the $X_{mono}$ states in a monolayer TMDs ribbon with a finite size $L = 15\lambda$ along $y$ direction (see Appendix B for calculation details), we confirm the existence of helical edge states with energies located in the gap between the two lowest bands, see Fig. 5(b). In our calculation we set $y \in [-0.1\lambda, 14.9\lambda]$ which leads to different electrostatic potentials at the two edges. The edge states at opposite positions therefore have distinct dispersions. Without the e-h exchange interaction, edge states in opposite valleys are decoupled and exhibit counterpropagating directions, with propagation directions reversed after switching the edge position. Fig. 5(c) is the 2D map of $\Delta_{12}$ in $(t, \lambda)$ parameter space. The calculated $\mathcal{C}_1$ values of different $\Delta_{12} > 0$ regimes give a topological phase diagram for the **K** valley $X_{mono}$.

For bright $X_{mono}$ in the spin-singlet and intravalley configuration, the e-h exchange interaction efficiently couples **K** and $\bar{\mathbf{K}}$ valleys [19,34-36]. The strength of the exchange interaction is proportional to the exciton oscillator strength, thus is finite only for s-type Rydberg states. For the considered 2s and $2p_\pm$ states, the e-h exchange Hamiltonian is

$$\begin{aligned}\hat{H}_{ex} = &\sum_{\mathbf{k}} |\mathbf{k}, 2s\rangle\langle\mathbf{k}, 2s| \otimes |\mathcal{J}_{2s,\mathbf{k}}| (|+\rangle\langle+| + |-\rangle\langle-|) \\ &+ \sum_{\mathbf{k}} |\mathbf{k}, 2s\rangle\langle\mathbf{k}, 2s| \otimes (\mathcal{J}_{2s,\mathbf{k}}|+\rangle\langle-| + \mathcal{J}^*_{2s,\mathbf{k}}|-\rangle\langle+|).\end{aligned} \tag{3}$$

Here $|+\rangle$ ($|-\rangle$) represents **K** ($\bar{\mathbf{K}}$) valley, $\mathcal{J}_{2s,\mathbf{k}} \approx J_{2s} V(k) k^2 e^{-2i\theta_{\mathbf{k}}}$ is the exchange strength, $V(k) = \frac{1}{k(1+r_0 k)}$ is the **k**-space form of the Rytova-Keldysh Coulomb potential with $r_0$ the screening length of monolayer TMDs [63], and $J_{2s}$ is a constant. Below we set $J_{2s} = 0.05$ eV $\cdot$Å, $r_0 = 1$ nm and take other parameters the same as those in Fig. 5(a) for a better comparison. The obtained $X_{mono}$ bands can be separated into a series of isolated subsets, each formed by two nearly degenerate bands, see Fig. 5(d). The topological structure of each subset can be characterized by the spin Chern number $\mathcal{C}_s$ [64-66] (see Appendix C for calculation details), with $\mathcal{C}_s = 1$ for the lowest-energy subset in Fig. 5(d). The corresponding in-gap edge states in a ribbon with $y \in [-0.1\lambda, 14.9\lambda]$ are shown in Fig. 5(e). Compared to Fig. 5(b) where the edge states are gapless, the finite exchange interaction opens a gap at $k_x = 0$ and $\frac{\pi}{\sqrt{3}\lambda}$ in Fig. 5(e) due to the strong hybridization between **K** and $\bar{\mathbf{K}}$ valley edge states, where velocities and valley polarizations vanish. However, at momentums away from $k_x = 0$ and $\frac{\pi}{\sqrt{3}\lambda}$, edge states again show finite valley polarizations and counterpropagating directions. In order for the optically active edge states at $k_x = 0$ to be chiral, one can apply a magnetic field to introduce a Zeeman splitting between the two valleys and suppress their mixing. In Appendix B we show the calculated edge states under a small Zeeman splitting, which become fully valley polarized and exhibit finite velocities at $k_x = 0$.

A well-defined spin Chern number requires the corresponding band subset to be isolated from others by finite global gaps. The gap $\Delta_{12}$ between the lowest-energy subset and the second-lowest one is found to depend sensitively on $J_{2s}$. We show our calculated $\Delta_{12}$ as a function of $(J_{2s}, t)$ in Figs. 5(f), indicating that a nontrivial topological phase requires a weak $J_{2s}$ plus a moderate $t$ value. For $J_{2s}$ above a threshold or $t$ smaller than some critical value, $X_{mono}$ becomes gapless and its spin Chern number cannot be defined. Meanwhile, the system enters a trivial gapped phase with $\mathcal{C}_s = 0$ when $t$ becomes too large.

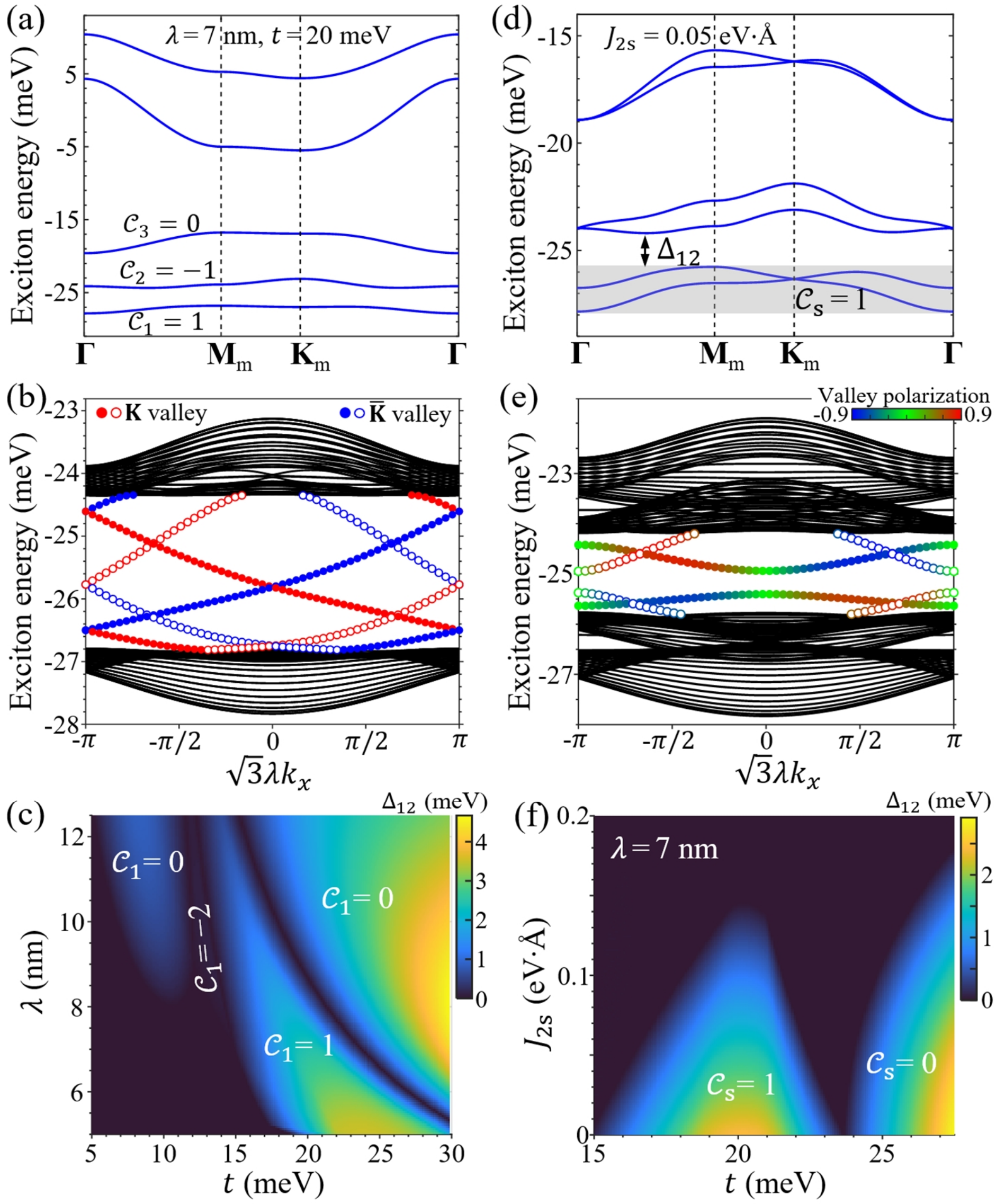

Figure 5. (a) **K** valley $X_{mono}$ bands under $\lambda = 7$ nm and $t = 20$ meV, with Chern numbers $(\mathcal{C}_1, \mathcal{C}_2, \mathcal{C}_3) = (1, -1, 0)$ for the three lowest bands. (b) 1D $X_{mono}$ bands in a TMDs ribbon with $y \in [-0.1\lambda, 14.9\lambda]$. Black lines are bulk states. Dots with red/blue color represent **K**/$\bar{\mathbf{K}}$ valley edge states. Filled and open dots represent edge states located near $y = -0.1\lambda$ and $14.9\lambda$, respectively. (c) A topological phase diagram for the **K** valley $X_{mono}$ in $(\lambda, t)$ parameter space. The color gives the global gap $\Delta_{12}$. (d) $X_{mono}$ bands under the effect of an e-h exchange interaction with $J_{2s} = 0.05$ eV·Å. The other parameters are identical to those in (a). The topological structure of the two nearly degenerate lowest bands is characterized by a spin Chern number $\mathcal{C}_s = 1$. (e) 1D $X_{mono}$ bands in a TMDs ribbon with $y \in [-0.1\lambda, 14.9\lambda]$. Black lines correspond to bulk states, while dots denote in-gap edge states with colors representing valley polarizations. Filled or open dots represent whether the edge states are located near $y = -0.1\lambda$ or $14.9\lambda$. (f) The global gap $\Delta_{12}$ shown in (d) as a function of $(J_{2s}, t)$ under $\lambda = 7$ nm.

## V. Discussion

Our work suggests a novel scheme for engineering topological exciton structures through manipulating the Rydberg DoF, which can be realized by a remotely imprinted periodic electrostatic potential. Such a scheme doesn't rely on spin-, valley- and layer-hybridizations, and can apply to excitons in various configurations. We found that the resultant topological lowest $X_{inter}$ band can be well suited for realizing the bosonic FCI. For $X_{mono}$ with large energy differences between 1s and $2p_{\pm}$, we expect the lowest band to be dominated by 1s state with trivial topologies when the potential is too weak to effectively hybridize 1s and $2p_{\pm}$. On the other hand, under an inversion-asymmetric electrostatic potential [67] with a strong enough strength, the lowest $X_{mono}$ band can also be topologically nontrivial, see the results in Appendix D. The finite out-of-plane electric dipole of $X_{inter}$ can facilitate the transport measurement of its topological states. Meanwhile, the edge states of $X_{mono}$ are found to exhibit in-plane electric dipoles with magnitudes ~ 1 $e\cdot$nm (see Appendix B), which originate from the $\hat{C}_3$ symmetry breaking at the edges.

The finite splitting $E_{2p_+} - E_{2p_-}$ proportional to the intrinsic Bloch band Berry curvature [48-50] plays an essential role for breaking the inversion symmetry. 2D materials with larger Berry curvatures (e.g., gapped multilayer graphene [68]) are expected to exhibit enhanced $E_{2p_+} - E_{2p_-}$ values, which can increase the global gap thus facilitate the realization of topological exciton states in experiments. A similar mechanism also applies to electrons and holes trapped at the 2D potential minimum, where the originally degenerate first-excited states with $p_{\pm}$ symmetries are split by the intrinsic Berry curvature. This implies that carriers in monolayer TMDs without involving the layer-pseudospin precession can also be topologically nontrivial when under a periodic electrostatic potential [69-71].

*Note added.* When finalizing the manuscript, we became aware of recent related works discussing the excitonic Chern insulators in TMDs moiré patterns [72,73]. The topological exciton structures proposed in these works originate from the spatially varying layer hybridization of the hole constituent. On the other hand, our proposal offers a completely different scheme, where it is the momentum-dependent hybridization between Rydberg states that gives rise to nontrivial topologies of the exciton.

## Acknowledgements

H.Y. acknowledges support by NSFC under grant No. 12274477. W.Y. is supported by the National Key R&D Program of China (2020YFA0309600), the Research Grant Council of Hong Kong (AoE/P-701/20, HKU SRFS2122-7S05), and New Cornerstone Science Foundation.

## Appendix A: Inter-Rydberg coupling strengths induced by a periodic electrostatic potential

In the TMDs layer where the exciton resides, a periodic electrostatic potential can be generated by spatially modulating charge distributions located in adjacent layers. Such spatial charge modulations intrinsically occur in ferroelectric domains of moiré-patterned hBN or TMDs bilayers without doping, or can be introduced by doping moiré-patterned multiplayer graphene or TMDs. Here we focus on the latter system, where the generated periodic electrostatic potential in the exciton-layer can be simply tuned through varying the doping density in the charge-layer. In TBG, the doped carriers accumulate at AA-stacked regions which form a triangular lattice; meanwhile in TMDs moiré patterns, correlated insulators with various lattice types can form under certain integer and fractional fillings. The generated periodic electrostatic potential in the exciton-layer can be Fourier expanded as $U(\mathbf{r}) = \sum_{\mathbf{G}} e^{i\mathbf{G}\cdot\mathbf{r}} U(\mathbf{G})$, with $\mathbf{G}$ the reciprocal lattice vector of the moiré-patterned charge modulation. $|U(\mathbf{G})|$ decays rapidly with $G$, therefore it is sufficient to keep the nonzero moiré reciprocal lattice vectors with the smallest magnitude. For a triangular-type charge distribution, such reciprocal lattice vectors correspond to $\mathbf{G} = \pm\mathbf{b}_{1,2,3}$ with $U(\mathbf{b}_1) = U(-\mathbf{b}_1) = U(\mathbf{b}_2) = U(-\mathbf{b}_2) = U(\mathbf{b}_3) = U(-\mathbf{b}_3) \equiv U_0$. Fig. 1(b) shows the corresponding landscape of $U(\mathbf{r})$ which has a modulation range of $9U_0$.

$U(\mathbf{r})$ introduces coupling strengths $t_{2\mathrm{p}_\pm,n\mathrm{s}}(\mathbf{G}) \equiv iU(\mathbf{G})\mathbf{G}\cdot\langle \mathbf{2p}_\pm|\hat{\mathbf{r}}|n\mathbf{s}\rangle = \frac{i}{\sqrt{2}}U(\mathbf{G})G|\langle \mathbf{2p}_\pm|\hat{\mathbf{r}}|n\mathbf{s}\rangle|e^{\mp i\theta_{\mathbf{G}}}$ and $t_{2\mathrm{p}_\pm,3\mathrm{d}_\pm}(\mathbf{G}) \equiv iU(\mathbf{G})\mathbf{G}\cdot\langle \mathbf{2p}_\pm|\hat{\mathbf{r}}|\mathbf{3d}_\pm\rangle = \frac{i}{\sqrt{2}}U(\mathbf{G})G|\langle \mathbf{2p}_\pm|\hat{\mathbf{r}}|\mathbf{3d}_\pm\rangle|e^{\pm i\theta_{\mathbf{G}}}$ in the exciton-layer, with $\hat{\mathbf{r}}$ the e-h relative position operator and $\theta_{\mathbf{G}}$ the direction angle of $\mathbf{G}$. To evaluate its magnitude, we model the charge distribution as a series of Gaussian wave packets located at moiré lattice sites $\mathbf{R}_n$, with a total charge density $\rho(\mathbf{r}) = \frac{\nu}{\pi\sigma^2}\sum_n e^{-(\mathbf{r}-\mathbf{R}_n)^2/\sigma^2}$. Here $\sigma$ is the wave packet width, and $\nu$ stands for the filling factor (defined as the number of doped carriers in each moiré supercell). This leads to $U(\mathbf{G}) \approx \frac{4\pi}{\sqrt{3}\lambda}\frac{\nu}{\epsilon\lambda G}\frac{e^{-G^2\sigma^2/4-Gd}}{(1+r_0G)(1+r_0'G)-r_0r_0'G^2e^{-2Gd}}$ [53,63]. Here $\epsilon$ is the dielectric constant of the surrounding environment, $\lambda$ is the wavelength of the charge modulation pattern, $r_0$ and $r_0'$ are screening lengths of the exciton-layer and charge-layer, respectively, and $d \approx 0.6$ nm corresponds to their vertical distance. We assume the system is encapsulated by thick hBN substrate and capping layer, this leads to $\epsilon = 4.5$ and $r_0 = 1$ nm for monolayer TMDs. Experiments have obtained $|\langle \mathbf{2p}_\pm|\hat{\mathbf{r}}|\mathbf{1s}\rangle| \approx 1$ nm in monolayer $MoSe_2$ [46], whereas numerical calculations give $|\langle \mathbf{2p}_\pm|\hat{\mathbf{r}}|\mathbf{2s}\rangle| \approx 4$ nm which is three to four times larger than $|\langle \mathbf{2p}_\pm|\hat{\mathbf{r}}|\mathbf{1s}\rangle|$ but slightly smaller than $|\langle \mathbf{2p}_\pm|\hat{\mathbf{r}}|\mathbf{3d}_\pm\rangle| \approx 5$ nm [59]. In TBG, the doped carriers accumulate at AA-stacked regions to form a triangular-type moiré pattern [51,52], whose distribution width $\sigma$ is expected to increase with $\nu$ due to the Pauli exclusion and Coulomb repulsion between carriers. Here we model it with a simple linear function $\sigma = (1 + 0.1\nu)$ nm. Fig. 6(a) shows the induced $t \equiv |t_{2\mathrm{p}_\pm,2\mathrm{s}}(\mathbf{b}_j)|$ as a function of $\nu$ under several $\lambda$ values, where we set $r_0' = 5$ nm considering the large screening effect of TBG. In TMDs moiré patterns, triangular-type Mott insulators are formed under integer fillings ($\nu = 1, 2, 3, \ldots$), and the corresponding $t$ values as functions of the wavelength $\lambda$ are displayed in Fig. 6(b) (setting $r_0' = r_0 = 1$ nm and $\sigma = (1 + 0.1\nu)$ nm). For both cases, $t$ falls is in the order of $O(10)$ meV and can be tuned through varying the doping density $\nu$ and wavelength $\lambda$.

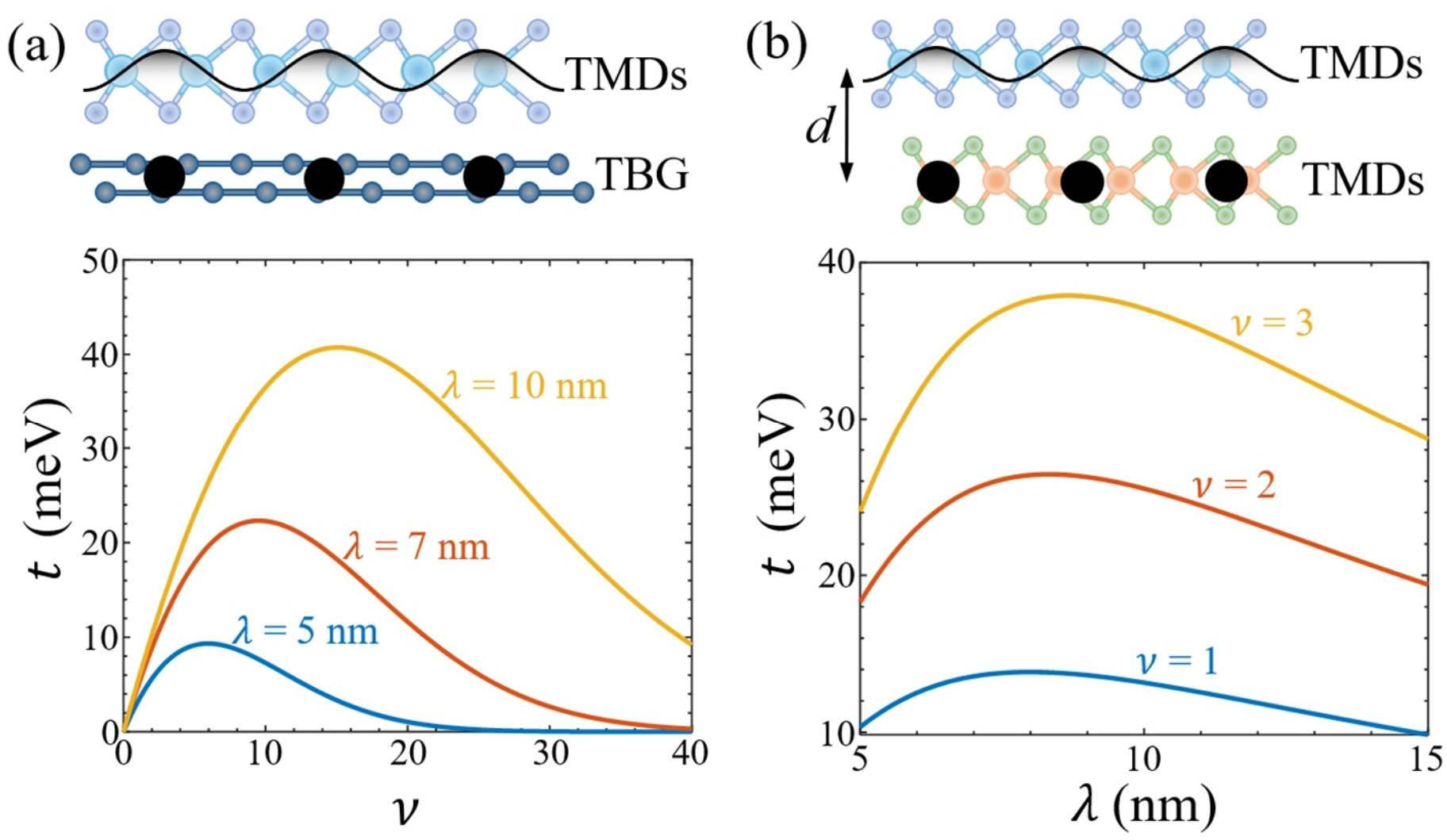

Figure 6. (a) Upper panel: a schematic illustration of a TMDs/TBG van der Waals structure, where excitons in monolayer TMDs experience a periodic electrostatic potential from charge distributions in TBG. Lower panel: the strength of $t$ as a function of the filling factor $\nu$ in TBG under three different wavelengths. (b) Upper panel: a schematic illustration of a bilayer TMDs moiré pattern, where a periodic electrostatic potential in the upper-layer is generated by charge distributions in the lower-layer. Lower panel: $t$ as a function of $\lambda$ under filling factors $\nu$ = 1, 2 and 3 in the lower-layer TMDs. We set $\epsilon$ = 4.5, $r_0$ = 1 nm, $d$ = 0.6 nm, $|\langle \mathbf{2p}_\pm|\hat{\mathbf{r}}|\mathbf{2s}\rangle|$ = 4 nm and $\sigma = (1 + 0.1\nu)$ nm in both (a) and (b), but $r_0'$ = 5 nm in (a) and $r_0'$ = 1 nm in (b).

## Appendix B: Solving the exciton edge states in a TMDs ribbon

We consider a system in a ribbon geometry with $X \in (-\infty, +\infty)$ and $Y \in [\xi, L+\xi]$, where $\mathbf{R} \equiv X\mathbf{x} + Y\mathbf{y}$ is the CoM coordinate for the exciton. For suitable choices of $\xi$ and $L$, the two edges located at $Y = \xi$ and $Y = L + \xi$ experience different electrostatic potentials, and the corresponding exciton edge states can be easily distinguished due to their distinct energies. The basis state for the exciton CoM motion is set as $\langle \mathbf{R}|k_x, l_y\rangle \equiv \sqrt{\frac{2}{A}} e^{ik_x X} \sin\left(\frac{l_y \pi (Y-\xi)}{L}\right)$, with $A$ the normalization area and $l_y$ = 1, 2, 3, …. After including the e-h relative wave function $|n\rangle = |1\mathrm{s}\rangle, |2\mathrm{p}_\pm\rangle, |2\mathrm{s}\rangle, \ldots$, the total basis state is $|k_x, l_y, n\rangle \equiv |k_x, l_y\rangle \otimes |n\rangle$ with $k_x \in \left[-\frac{\pi}{\sqrt{3}\lambda}, \frac{\pi}{\sqrt{3}\lambda}\right)$. The $\mathbf{K}$ valley exciton Hamiltonian can then be expressed as $\hat{H} = \sum_{k_x} (\hat{H}_{1,k_x} + \hat{H}_{2,k_x})$, with

$$\begin{aligned} \hat{H}_{1,k_x} &= \sum_{G_x} \sum_{l_y n} \left[\frac{\hbar^2}{2M}\left((k_x + G_x)^2 + \frac{l_y^2 \pi^2}{L^2}\right) + E_n\right] |k_x + G_x, l_y, n\rangle\langle k_x + G_x, l_y, n|, \\ \hat{H}_{2,k_x} &= 2 \sum_{\mathbf{G} \neq 0} \sum_{l_y l_y' n n'} F_{l_y, l_y'}^{n,n'}(\mathbf{G}) \sum_{G_x'} |k_x + G_x' + G_x, l_y, n\rangle\langle k_x + G_x', l_y', n'|. \end{aligned} \tag{A1}$$

$\hat{H}_{1,k_x}$ is the free exciton Hamiltonian, $\hat{H}_{2,k_x}$ corresponds to the effect of the periodic electrostatic potential with $F_{l_y, l_y'}^{n,n'}(\mathbf{G}) \equiv iU(\mathbf{G})\mathbf{G} \cdot \langle n|\hat{\mathbf{r}}|n'\rangle \int_0^L dY \sin\left(\frac{l_y' \pi Y}{L}\right) \sin\left(\frac{l_y \pi Y}{L}\right) \frac{e^{iG_y(Y+\xi)}}{L}$ and $\mathbf{G} \equiv G_x \mathbf{x} + G_y \mathbf{y}$.

The exciton edge states can be solved from $\hat{H}_{1,k_x} + \hat{H}_{2,k_x}$. Fig. 7(a) shows the 1D bands of $\mathbf{K}$ valley $X_{\mathrm{mono}}$ in a TMDs ribbon with $Y \in [-0.1\lambda, 14.9\lambda]$, where bulk and edge states are shown as lines and dots, respectively. Other parameters are the same as Fig. 5(b) in the maintext. We focus on the edge state $\left|\psi_{k_x=0}^{(\text{edge})}\right\rangle$ localized near the $y=-0.1\lambda$ edge with $k_x=0$, whose spatial distribution of the wave function intensity is shown in Fig. 7(b). As the edge breaks the $\hat{C}_3$ symmetry, the hybridization between $|2s\rangle$ and $|2p_{\pm}\rangle$ introduces a finite in-plane electric dipole to $\left|\psi_{k_x=0}^{(\text{edge})}\right\rangle$. The corresponding in-plane electric dipole as a function of **R** is shown in Fig. 7(c), whose average value $\mathbf{D}_{k_x=0}^{(\text{edge})}\equiv -e\left\langle\psi_{k_x=0}^{(\text{edge})}\middle|\hat{\mathbf{r}}\middle|\psi_{k_x=0}^{(\text{edge})}\right\rangle$ is perpendicular to the edge (along *y* axis). For general values of $k_x$, the corresponding in-plane electric dipoles $-e\left\langle\psi_{k_x}^{(\text{edge})}\middle|\hat{\mathbf{r}}\middle|\psi_{k_x}^{(\text{edge})}\right\rangle$ are also finite and along *y* axis, whose values are indicated as dot colors in Fig. 7(a).

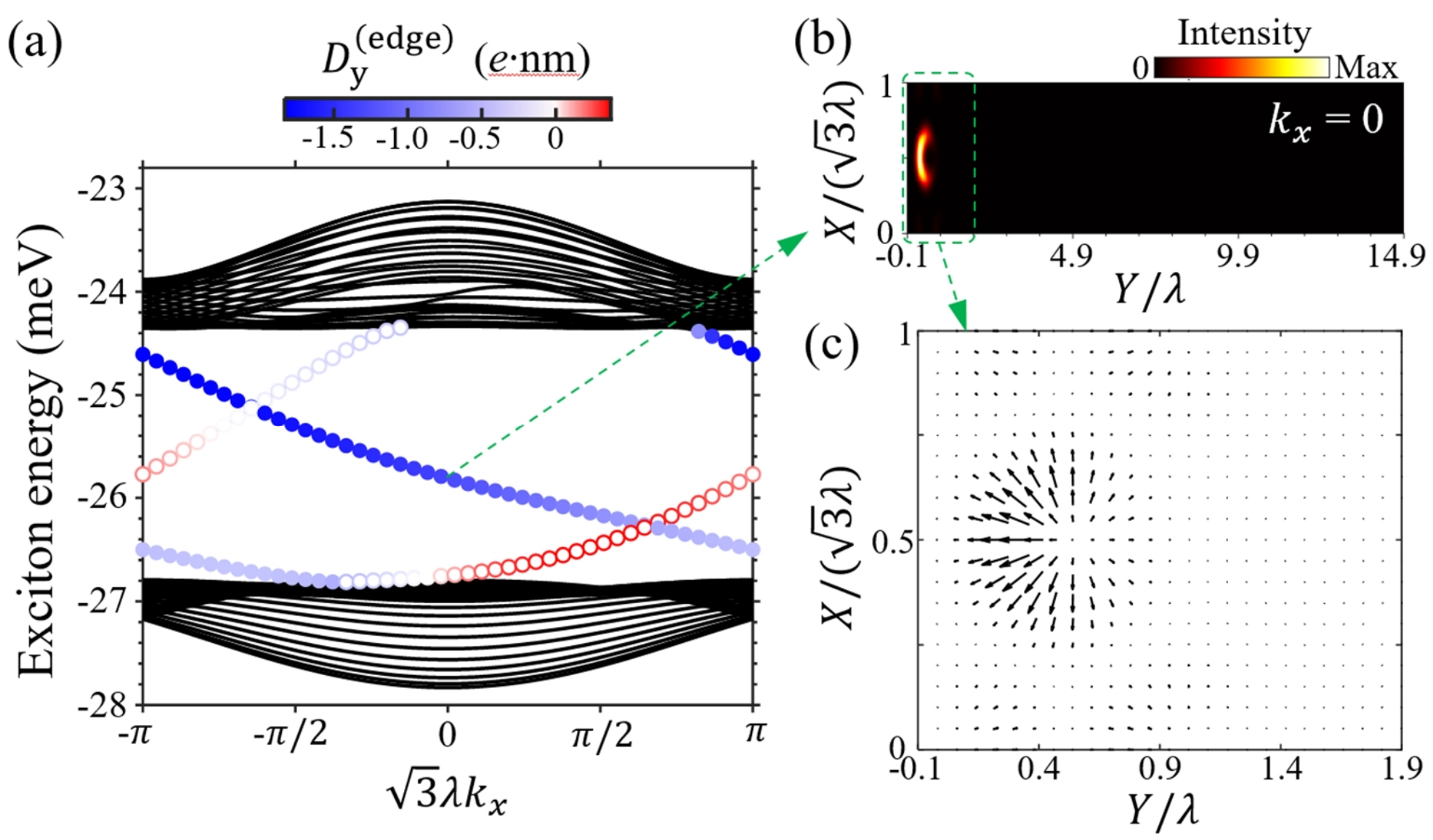

Figure 7. (a) **K** valley $X_{\text{mono}}$ bands in a TMDs ribbon with $y\in[-0.1\lambda, 14.9\lambda]$, where $\lambda=7$ nm is the moiré wavelength. Black lines and colored dots correspond to bulk and edge states, respectively. Filled (open) dots represent edge states located near $y=-0.1\lambda$ ($y=14.9\lambda$), with the color indicating the average in-plane electric dipole strength. (b) The spatial distribution of the CoM wavefunction intensity for the edge state at $k_x=0$, which are localized near $y=-0.1\lambda$. (c) The spatial distribution of the in-plane electric dipole of the edge state at $k_x=0$.

Eq. (A1) above does not take into account the valley-dependent e-h exchange interaction. The basis state becomes $|k_x, l_y, n\rangle\otimes|\pm\rangle$ when the valley pseudospin of the exciton is included, with $|+\rangle$ ($|-\rangle$) denoting **K** ($\bar{\mathbf{K}}$) valley. Following Ref. [41], we write the e-h exchange interaction Hamiltonian as $\hat{H}_{\text{ex}}=\sum_{k_x}\hat{H}_{\text{ex},k_x}$ with

$$\begin{aligned}
\hat{H}_{\text{ex},k_x}=&\sum_{G_x}\sum_{l_y n}|k_x+G_x,l_y,n\rangle\langle k_x+G_x,l_y,n|\otimes|J_{n,\boldsymbol{k}}|(|+\rangle\langle+|+|-\rangle\langle-|)\\
&+\sum_{G_x}\sum_{l_y n}|k_x+G_x,l_y,n\rangle\langle k_x+G_x,l_y,n|\otimes\mathrm{Re}(J_{n,\boldsymbol{k}})(|+\rangle\langle-|+|-\rangle\langle+|)\\
&+\sum_{G_x}\sum_{l_y\neq l_y'}\sum_{n}\frac{k_x+G_x}{L}\left(\frac{|J_{n,\boldsymbol{k}}|}{|\boldsymbol{k}|^2}+\frac{|J_{n,\boldsymbol{k}'}|}{|\boldsymbol{k}'|^2}\right)f_{l_y,l_y'}|k_x+G_x,l_y,n\rangle\langle k_x+G_x,l_y',n|\\
&\otimes(-i|+\rangle\langle-|+i|-\rangle\langle+|).
\end{aligned}\tag{A2}$$

Here $\boldsymbol{k} \equiv (k_x + G_x)\mathbf{x} + \frac{l_y \pi}{L}\mathbf{y}$, $\boldsymbol{k}' \equiv (k_x + G_x)\mathbf{x} + \frac{l'_y \pi}{L}\mathbf{y}$, $J_{n,\boldsymbol{k}} \equiv \frac{J_n|\boldsymbol{k}|}{1+r_0|\boldsymbol{k}|} e^{-2i\theta_{\boldsymbol{k}}}$ and $f_{l_y,l'_y} = -i\frac{2l'_y\pi}{L}\int_0^L dY \sin\left(\frac{l_y\pi Y}{L}\right)\cos\left(\frac{l'_y\pi Y}{L}\right) = -i\frac{2l_y l'_y}{l_y^2 - l'^2_y}\left[1-(-1)^{l_y+l'_y}\right]$. Note that $J_n$ is finite only for s-type Rydberg states. The edge states under the e-h exchange interaction can then be solved from $\widehat{H}_{1,k_x} + \widehat{H}_{2,k_x} + \widehat{H}_{\mathrm{ex},k_x}$.

An out-of-plane magnetic field introduces a Zeeman splitting $\Delta E_\mathrm{B}$ to excitons in $\mathbf{K}$ and $\bar{\mathbf{K}}$ valleys, which can modify the energies and valley polarizations of the edge states for bright $X_{mono}$ in the spin-singlet and intravalley configuration. Fig. 8(a) and 8(b) show the resultant $X_{mono}$ bands in a 2D layer and 1D ribbon, respectively, when a small Zeeman splitting $\Delta E_\mathrm{B}$ = 0.4 meV is introduced to $X_{mono}$ in Fig. 5(a,b). The optically bright edge states at $k_x = 0$ now exhibit large valley polarizations and finite group velocities (Fig. 8(b)), resulting in helical edge states that can be optically accessed. Fig. 8(c,d) correspond to results under a relatively large Zeeman splitting $\Delta E_\mathrm{B}$ = 3 meV. The lowest and fourth-lowest bands are now separated from others by global gaps, with Chern numbers $\mathcal{C}_1 = \mathcal{C}_4 = -1$. The edge states shown in Fig. 8(d) are now fully valley polarized, with the valley indices and propagation directions locked to the energy positions.

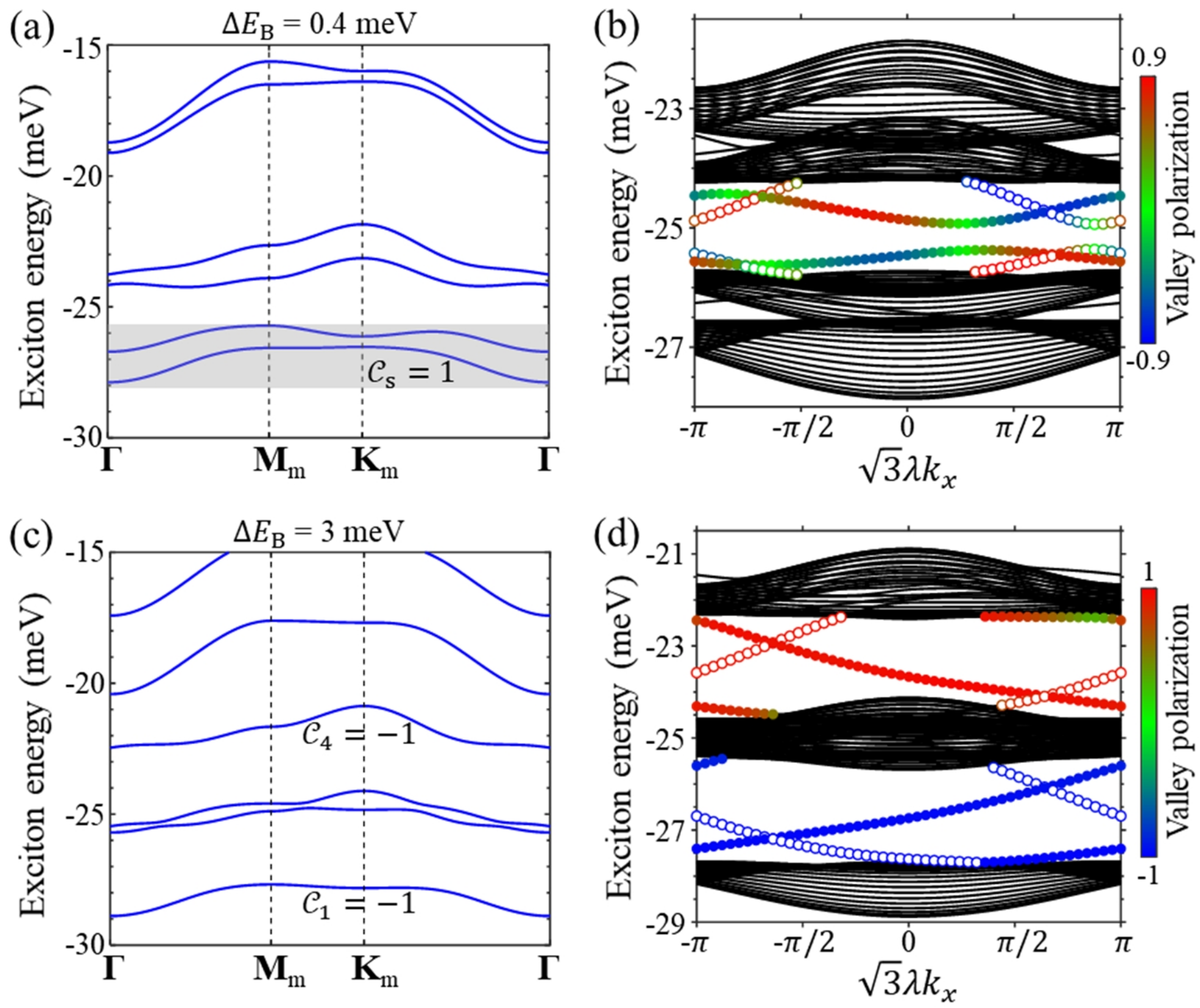

Figure 8. (a) 2D exciton bands under a Zeeman splitting $\Delta E_\mathrm{B}$ = 0.4 meV, with the other parameters the same to those in Fig. 5(a,b). The spin Chern number of the two nearly degenerate lowest bands is $\mathcal{C}_\mathrm{S}$ = 1. (b) The corresponding 1D exciton bands in a TMDs ribbon. (c) and (d) Results under $\Delta E_\mathrm{B}$ = 3 meV. The Chern numbers of the lowest band and fourth band are $\mathcal{C}_1 = \mathcal{C}_4 = -1$.

## Appendix C: Calculation of the spin Chern number

The spin Chern number introduced by Sheng et al [64,66], was later redefined in Ref. [65] as the half difference between the Chern numbers of the occupied space's (pseudo) spin-up and spin-down sectors, without involving any boundary conditions. Following Ref. [65], the subspace of the occupied bands can be partitioned into two nontrivial spin sectors by using the projected pseudospin operator $\hat{P}(\mathbf{k})\hat{\sigma}_z\hat{P}(\mathbf{k})$, where $\hat{P}(\mathbf{k})$ is the projection operator onto the occupied subspace with CoM momentum $\mathbf{k}$, and and $\hat{\sigma}_z$ is the Pauli matrix for the spin/valley. If amplitude of the exchange interaction $J_{2s}$ does not exceed a threshold value, the spectrum of $\hat{P}(\mathbf{k})\hat{\sigma}_z\hat{P}(\mathbf{k})$ consists of two isolated eigenvalues with eigenstates $\psi_{\pm}(\mathbf{k})$ for all $\mathbf{k}$ of the Brillouin torus. The spin Chern number $\mathcal{C}_s$ is given by $\mathcal{C}_s = \frac{1}{2}(\mathcal{C}_+ - \mathcal{C}_-)$ [65], where $\mathcal{C}_{\pm}$ are the Chern number of $\psi_{\pm}$. In a word, though the Chern number of the entire system is zero, we can use the operator $\hat{P}(\mathbf{k})\hat{\sigma}_z\hat{P}(\mathbf{k})$, to slash it into two nontrivial parts. We calculate the spin Chern number $\mathcal{C}_s$ numerically and use this integer topological invariant to distinguish different topological phases in our model.

## Appendix D: Emergence of topological monolayer excitons near the 1s energy

An inversion asymmetric electrostatic potential can emerge in monolayer TMDs adjacent to a TBG/hBN double-moiré structure, where the TBG moiré is commensurate with the graphene/hBN moiré but with distinct wavelengths [67]. Fig. 9(a) shows such a double-moiré pattern where A and B correspond to inequivalent AA sites in TBG aligned with different locales in the graphene/hBN moiré. These two sub-sites have different charge densities under a finite doping, whose ratio $\delta \in [0,1]$ is treated as a tunable parameter below. Now $t_{nn'}(\mathbf{b}_j)$ in maintext Eq. (2) should be replaced by $t_{nn'}(\mathbf{b}_j)(1 + \delta e^{-i\mathbf{b}_j\cdot\mathbf{R}_{AB}})$.

Below we consider 1s, 2s and 2p$_{\pm}$ Rydberg states for **K** valley X$_{\text{mono}}$ under such a periodic electrostatic potential, without including the e-h exchange interaction. We set $E_{1\text{s}} = 0$, $E_{2\text{p}_-} = 105$ meV, $E_{2\text{p}_+} = 120$ meV, $E_{2\text{s}} = 165$ meV, $t = 4t'$. Fig. 9(b) gives the calculated global gap value $\Delta_{12}$ as a function of $(\delta, t)$ under $\lambda = 5$ nm. The lowest band becomes topologically nontrivial with a Chern number $\mathcal{C}_1 = 1$ in a parameter regime with $\delta \approx 0.5$ and large $t$. Fig. 9(c) shows the three lowest bands of **K** valley X$_{\text{mono}}$ under $\delta = 0.5$, $\lambda = 5$ nm and $t = 60$ meV, which have Chern numbers $(\mathcal{C}_1, \mathcal{C}_2, \mathcal{C}_3) = (1, 0, -1)$. The corresponding edge states in a monolayer ribbon with $y \in [-0.1\lambda, 19.9\lambda]$ are given in Fig. 9(d), which again exhibit unidirectional propagations when the energies are in the gap.

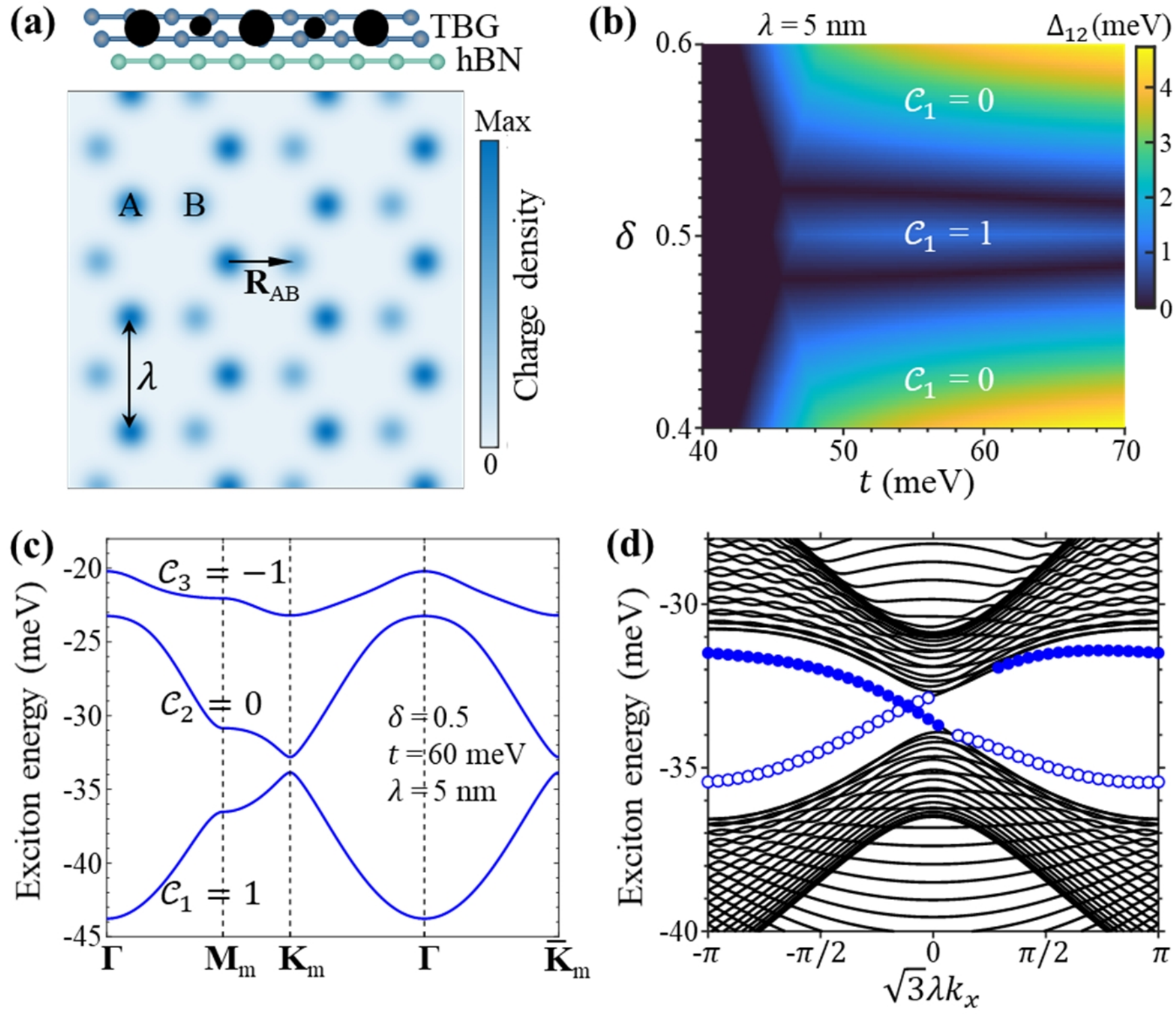

Figure 9. (a) A schematic illustration of the inversion asymmetric honeycomb-type charge density distribution, which can be realized in a TBG/hBN double moiré structure. (b) The global gap value $\Delta_{12}$ of $X_{mono}$ as a function of $(\delta, t)$ under $\lambda = 5$ nm, $E_{1s} = 0$, $E_{2p_-} = 105$ meV, $E_{2p_+} = 120$ meV, $E_{2s} = 165$ meV. (c) The three lowest bands of the **K** valley $X_{mono}$ under $\lambda = 5$ nm, $\delta = 0.5$ and $t = 4t' = 60$ meV, with Chern numbers $(\mathcal{C}_1, \mathcal{C}_2, \mathcal{C}_3) = (1, 0, -1)$. (d) 1D $X_{mono}$ bands in a monolayer TMDs ribbon with $y \in [-0.1\lambda, 19.9\lambda]$. Black lines depict the bulk bands. Filled and open dots represent edge states located near $y = -0.1\lambda$ and $y = 19.9\lambda$ edges, respectively.